\documentstyle[12pt]{article}
\normalbaselines
\catcode `\@=11
\@addtoreset{equation}{section}

\def\section{\@startsection {section}{1}{\z@}{-3.5ex plus -1ex minus
     -.2ex}{2.3ex plus .2ex}{\normalsize\bf}}
\def\subsection{\@startsection{subsection}{2}{\z@}{-3.25ex plus -1ex minus
 -.2ex}{1.5ex plus .2ex}{\normalsize\bf}}

\def\thebibliography#1{\section*{References\markboth
  {REFERENCES}{REFERENCES}}\list
  {[\arabic{enumi}]}{\settowidth\labelwidth{[#1]}\leftmargin\labelwidth
  \advance\leftmargin\labelsep
  \usecounter{enumi}}
  \def\newblock{\hskip .11em plus .33em minus -.07em}
  \sloppy
  \sfcode`\.=1000\relax}
 
\catcode `\@=12

\begin{document}

\vspace*{2.4cm}
\noindent
{ \bf A DYNAMICAL MECHANISM FOR THE SELECTION OF PHYSICAL STATES
      IN `GEOMETRIC QUANTIZATION SCHEMES'}\vspace{1.3cm}\\
\noindent
\hspace*{1in}
\begin{minipage}{13cm}
P. Maraner \vspace{0.3cm}\\
\makebox[3mm]Center for Theoretical Physics,\\
\makebox[3mm]Laboratory for Nuclear Science and INFN,\\ 
\makebox[3mm]Massachusetts Institute of Technology, \\
\makebox[3mm]Cambridge, MA 02139-4307, USA
\end{minipage}

\vspace*{0.5cm}

\begin{abstract}
\noindent 
Geometric quantization procedures go
usually through an extension of the original theory (pre-quantization) 
and  a subsequent reduction (selection of the physical states). 
In this context we describe a full geometrical mechanism which 
provides dynamically the desired reduction.

\end{abstract}

\section[]{\hspace{-4mm}.\hspace{2mm} 
               THE STANDARD VIEWPOINT ON QUANTIZATION:\\ 
               STATES, OBSERVABLES AND TIME EVOLUTION}

\hspace*{0.8cm} The usual way to think of a physical system proceeds
in two steep. First: {\sl kinematics}, that is the specification of 
the possible states and of the observable quantities. Second: 
{\sl dynamics}, that is  the description of the time evolution of the 
representative point of the system over the space of all the 
possible states. As a very typical example we may think of Hamiltonian 
mechanics, where the states of a system with $n-$degrees of freedom are
specified by the canonical variables $q^\mu,p_\mu$, $\mu=1,...,n$, while 
observables are identified with smooth 
functions of $q$ and $p$. Dynamics is obtained by pointing out 
a privileged observable, namely the energy of the system $h(q,p)$, 
by means of the canonical flow generated on the phase space. 
 Let us note that still at this classical level, it turns out that a 
very few of all the possible observables of the theory play a concrete 
role in the physical description of the system. Furthermore, the selection 
of relevant observables goes typically through dynamical considerations,
making the sharp separation of kinematics and dynamics  an artificial one.

 Nevertheless the standard way to look at quantization moves from 
this viewpoint seeking a correspondence between the formal structures
of classical and quantum mechanics: {\em states}, {\em observables}
and {\em time evolution}. In order to make the  quantization procedure 
a sensible one, that is capable to reproduce standard quantum mechanics, 
four conditions are usually required: ($Q1$) the correspondence is 
asked to be linear; ($Q2$) the constant function $1$ has to be mapped 
on to the identity operator; ($Q3$) the Poisson brackets should become $i$ 
times the commutators; and ($Q4$) the canonical variables $q$ and $p$ 
should act irreducibly on the quantum Hilbert space.
 There are of course many critiques that can be moved to such an approach.
As a matter of fact, it results that it is impossible to quantize the 
whole algebra of classical observables without violating at least one of 
these conditions. The quantization program fails already at the kinematical
level. Thought different viewpoints have been expressed in the literature 
it is the common believe that the 
weak point of the above construction is that of requiring
that such only locally defined objects as the canonical coordinates
have to be promoted to globally well defined operators acting 
irreducibly on the quantum Hilbert space. 

 Once condition ($Q4$) is lifted, it is infact possible to 
proceed in a very general and elegant manner to the construction of 
the desired correspondence: Kostant's {\em pre-quantization}
scheme \cite{Wo80}. The whole huge algebra of smooth functions on the 
phase space
is mapped in the algebra of formally self-adjoint operators on a
suitable Hilbert space in such a way that conditions ($Q1$),
($Q2$) and ($Q3$) are fulfilled. The problem is
that the pre-quantum Hilbert space is too large for physics. 
 In maintaining such an approach it is therefore necessary to introduce
a mechanism capable of selecting the subspace of physical states.
 Thought many different approaches have been suggested, the 
general viewpoint is  that of picking out a {\em real} or 
{\em complex polarization} on the classical phase space and requiring
that the physical states are the one preserving the polarization. 
Such a  prescription---definitely of kinematical character---works well
as long as the quantization of systems with a high degree of 
symmetry is concerned \cite{On76}, but it appears more and more problematic
as soon as the dynamics of systems with less symmetry or no symmetry 
at all is considered. There is in fact no longer guarantee that time
evolution respects the polarization, and physical states may evolve 
in non physical ones. 

 It is the aim of this paper to present a slightly different approach
to the problem focusing more on the dynamical aspects rather than on
the kinematical ones \cite{KM97}. Working in a coordinate free manner 
we will be facing the problem of directly defining the quantum dynamics 
without going through the quantization of the whole algebra of classical 
observables. This yields a dynamical mechanism that produces the selection 
of the right set of physical quantum states.

\section{\hspace{-4mm}.\hspace{2mm} COORDINATE FREE QUANTIZATION}

\hspace*{0.8cm} A sensible quantization scheme should not depend 
on the choice of coordinates. Before discussing quantization let 
us therefore  briefly recall how is possible to formulate Hamiltonian 
mechanics in a coordinate free manner.

{\it Coordinate Free Formulation of Hamiltonian Mechanics} \cite{HM}: 
It is convenient to denote phase space coordinates by means
of a single variable $\xi=(q^1,...,q^n,p_1,...,p_n)$. In this canonical 
coordinate frame we introduce the skew-symmetric two-tensors 
$\omega_{ij}$, $i=1,...,2n$,
\begin{equation}
\omega_{ij}=\pmatrix{0 & -I \cr
                     I &  0 },
\label{sf}
\end{equation}
$I$ is the $n$-dimensional identity matrix, 
and ${\bar\omega}^{ij}$ defined 
by the relation $\omega_{ik}{\bar\omega}^{kj} =\delta_i^j$.
The fundamental Poisson bracket may so be recasted in the covariant form 
$\{\xi^i,\xi^j\}={\bar\omega}^{ji}$. We note that a canonical transformation 
does not affect the form of $\omega_{ij}$. Furthermore, the information on
the canonical structure being contained in $\omega_{ij}$, we are now free to 
introduce arbitrary coordinate frames, not necessarily preserving \ref{sf}. 
 The phase space of a Hamiltonian system may so be identified with a 
{\sl symplectic manifold}, that is a $2n$-dimensional manifold ${\cal M}$
equipped with a closed nondegenerate two-form, the {\sl symplectic form}
$\omega_{ij}$. For a Lagrangian system with configuration space $Q$ 
the symplectic manifold ${\cal M}$ have to be identified with the cotangent 
bundle $T^\star\!Q$, but this is not the most general case. Many 
system of physical interest are not included in this class and the discussion
of more general phase spaces is necessary. 
 In order to give a coordinate free formulation of the dynamics of the 
system we have  to introduce the canonical one-form $\theta_i$, defined
by the relation $\omega_{ij}=\partial_i\theta_j-\partial_j\theta_i$. $\theta_i$
is defined up to the total derivative of an arbitrary phase space function,
$\theta_i\rightarrow\theta_i+\partial_i\chi$ (note the formal 
equivalence of the canonical one and two-forms with a vector potential 
and a magnetic field!). Dynamics is then defined by 
means of Hamilton's principle 
\begin{equation}
\delta\int(\theta_i{\dot\xi}^i- 
h(\xi))dt=0
\end{equation}

{\it The Geometrical Background of (pre-)Quantization} \cite{Wo80}: 
In a coordinate free language the problem of pre-quantization 
may therefore be recasted in the following terms. For every symplectic
manifold ${\cal M}$ construct a Hilbert Space $H({\cal M})$
such that it is possible to exhibit a map form the algebra of smooth
function on ${\cal M}$ into that of the formally self-adjoint operators
on $H({\cal M})$ satisfying conditions ($Q1$), ($Q2$) and ($Q3$).
 This is achieved by identifying $H({\cal M})$ with the Hilbert 
space of the square integrable sections of the line bundle $L$ on ${\cal M}$
having the symplectic two-form $\omega_{ij}$ as curvature form. The 
correspondence between classical and quantum observables may then
be constructed in terms of the covariant derivative on the line 
bundle. Without going too much into details we recall that for 
${\cal M}=T^\star\!Q$, in a canonical coordinate frame and fixed
the gauge $\theta=(0,...,0,-q^1,...,-q^n)$ the general rule yields
\begin{eqnarray}
p_\mu&\rightarrow& -i\hbar{\partial\over\partial q^\mu},       \label{qcin} \\
q^\mu&\rightarrow&\ i\hbar{\partial\over\partial p_\mu}+q^\mu, \label{pcin} 
\end{eqnarray}
whereas $H(T^\star\!Q)$ roughly corresponds with the space of square 
integrable functions of $q$ and $p$, $\psi(q,p)$. In this simple case the 
selection of the right set of physical states is  achieved by 
requiring the wave functions to be constant in the $p_\mu$ directions,
that is $\partial\psi/\partial p_\mu=0$ for $\mu=1,...,n$. This is
of course a very simple case. Nevertheless it somehow suggest
to think of the phase space ${\cal M}$ as a sort of configuration
space on which the operators  $-i\hbar{\partial\over\partial\xi}=
(-i\hbar{\partial\over \partial q^\mu}, -i\hbar{\partial\over\partial 
p_\mu})$  play the role of the canonical momenta conjugate to 
$\xi=(q^\mu,p_\mu)$. We note that operators \ref{qcin}
and \ref{pcin} may then be identified with the kinematical momenta 
of a charged particle moving on ${\cal M}$ in the magnetic field
$\omega_{ij}$ represented by the vector potential $\theta_i$.

\section[]{\hspace{-4mm}.\hspace{2mm} A SLIGHTLY DIFFERENT VIEWPOINT:\\
                    DYNAMICS AND QUANTIZATION ON $T^\star\!{\cal M}$}

\hspace*{0.8cm} A slightly different way to look at the pre-quantization 
scheme may infact be that of considering an extension of the  
mechanical system from the original phase space ${\cal M}$ 
to the enlarged phase space $T^\star\!{\cal M}$ (note: the cotangent 
bundle of the phase space!). Thought this introduces many ambiguities
it has the advantage that the quantization of a cotangent bundle is 
definitely simpler than the one of an arbitrary symplectic manifold. 
Ambiguities arise both in the extension and in the 
subsequent reduction of the system and many different mechanisms 
may be thought of to achieve the aim. From this perspective Kostant's 
pre-quantization represents one of the possible extension schemes 
and the selection of physical states by means of polarizations is 
only one of the possible reduction procedures. This way of facing
quantization appears as a promising one and has been adopted 
by many authors (see \cite{KlQ,Fe94,FL94,Go95,Jo96} for a few recent 
examples).
 In this context, we present a somehow peculiar quantization procedure 
constructed in the following way. We first extend classical dynamics 
from the phase space ${\cal M}$ to its cotangent bundle $T^\star\!{\cal M}$ 
by constructing a full geometrical theory depending on the parameter $\hbar$.
In the regime of small values of the parameter the theory reduces
dynamically to Hamiltonian mechanics. We then proceed to the quantization
on the extended theory and note how the same dynamical mechanism
provides the reduction to the physical sector of the quantum theory.

{\it A Full Geometrical Extension of Hamiltonian Mechanics}: 
We start with a Hamiltonian system with phase space ${\cal M}$ and 
Hamiltonian $h(\xi)$. Thought our theory is covariant in character it 
is useful to parameterize ${\cal M}$ by means of a canonical atlas, so 
that in every coordinate frame the symplectic structure $\omega_{ij}$ 
appears in the canonical form \ref{sf}. We now introduce a metric structure 
on ${\cal M}$ requiring that in every canonical frame the metric determinant
satisfies the condition $g(\xi)=h^{-2n}(\xi)$. We finally extend our 
mechanical system to $T^\star\!{\cal M}$ by defining dynamics by means 
of the variational principle 
\begin{equation}
\delta\int({1\over2}\hbar g_{ij}{\dot\xi}^i{\dot\xi}^j
                          +\theta_i{\dot\xi}^i)dt=0.
\label{gd}
\end{equation} 
$\hbar$ is Plank's constant over $2\pi$. We claim  that  
the phase space trajectories produced by \ref{gd} differ from the 
one produced by Hamilton's principle only over scales 
of order $\hbar$. Hamiltonian mechanics may therefore be regarded 
as the effective theory describing the small $\hbar$ regime of the 
geometrical theory \ref{gd}. Observe that although $\hbar$ 
appears into the theory, we are not claiming that \ref{gd}
describes quantum mechanics.  
We just find it very useful to incorporate Plank's constant
in the extension of the theory to $T^\star\!{\cal M}$
in such a way that this parameter controls dynamically 
the reduction of the extended theory to the original one.

 Before proceeding in the demonstration of our claim let us note 
that the variational principle \ref{gd} is formally equivalent
to that describing the free motion of a particle of mass $\hbar$
on the metric manifold ${\cal M}$ in the universal magnetic field 
represented by the symplectic form $\omega_{ij}$. 
It is therefore possible to visualize the mechanism responsible for
the reduction of our theory by thinking of a particle of mass $m$ 
and charge $e$ moving in a plane under the influence of a magnetic 
field of magnitude  $B$ normal to the plane. 
In the analogy the plane represents the 
phase space  of a one-dimensional system while the magnetic field its
symplectic structure. The regime of a small mass corresponds to that of a 
strong magnetic field, or equivalently, to that of a nearly homogeneous one. 
This problem, sometimes called the guiding center problem, has been 
extensively discussed in the literature
\cite{HGCM}. As long as the magnetic field 
may be considered as homogeneous the particle follows a circular 
orbit of  radius $r_m=mc|{\vec v}|/eB$ the center of which is 
motionless. For a very  small mass the circle is so narrow that 
the particle appears at rest. However, as soon as a weak inhomogeneity 
is introduced the center of the orbit---usually called {\em guiding
center}---starts drifting on the plane. Moreover, the guiding center 
motion is Hamiltonian.  
 We shall identify the guiding center motion with the motion of 
our original system while the rapid rotation around 
the effective trajectory with the degrees of freedom suppressed by
the reduction.

 Having this picture in mind we now sketch a formal demonstration.
Starting from the Lagrangian ${\cal L}(\xi,\dot\xi)={1\over2}\hbar g_{ij}
{\dot\xi}^i{\dot\xi}^j+\theta_i{\dot\xi}^i$ of the extended system
we proceed to the construction of the relative Hamiltonian formalism
by introducing the canonical momenta $p_i^\xi=\partial{\cal  L}/
\partial\dot\xi^i$ conjugate to the variables $\xi^i$. The Hamiltonian 
describing the dynamics of the extended system yields 
\begin{equation}
{\cal H}={1\over2\hbar}g^{ij}(\xi)(p_i^\xi-\theta_i)(p_j^\xi-\theta_j),
\label{ham1}
\end{equation}
where $g^{ij}$ denotes the inverse of the metric tensor. In order to discuss
the small $\hbar$ regime of the theory it is very convenient to replace
the set of canonical variable $\xi^i, p_i^\xi$, $i=1,...,2n$, with the
gauge covariant {\em kinematical momenta} and {\em guiding center coordinates}
$$
\Pi_i={1\over\hbar^{1/2}}(p_i^\xi-\theta_i) \hskip0.7cm\mbox{and}\hskip0.7cm
X^i =\xi^i+\hbar^{1/2}{\bar\omega}^{ij}\Pi_j.
$$
The new set of coordinates is canonical: $\Pi_\mu$ is conjugate to
$\Pi_{n+\mu}$ and $X^{n+\mu}$ to $X^\mu$, $\mu=1,...,n$. Furthermore,
in the new set of variables \ref{ham1} appears as the Hamiltonian 
of an $n$-dimensional harmonic oscillator with masses and frequencies
depending on the parameters $X^i$ and weakly on the `positions' and 
`velocities' $\Pi_i$. Since we are only interested in the small $\hbar$
regime of the theory it appears natural to expand $g^{ij}$ in powers of 
$\hbar^{1/2}$, 
\begin{equation}
{\cal H}={1\over2}g^{ij}(X)\Pi_i\Pi_j+{\cal O}(\hbar^{1/2})
\label{ham2}
\end{equation}   
which makes clear that only the dependence on the $X^i$ is relevant.
A further analysis of the commutation relation makes it clear that
the $X^i$ may be regarded as slow parameter of the system so 
that it is possible to perform a second canonical transformation
(see \cite{KM97} for details) bringing \ref{ham2} into the form
\begin{equation}
{\cal H}= h(X)\ {1\over2}\sum_i\Pi_i\Pi_i+{\cal O}(\hbar^{1/2}).
\label{ham3}
\end{equation}
The condition $g=h^{-2n}$ has been used. 
The guiding center motion described by the set of
canonical coordinates $X^i$ and the rapid rotation of the system 
around the guiding center trajectory are separated up to terms 
of order $\hbar^{1/2}$. Our demonstration is completed by 
observing that the radius of the circular phase space trajectory 
described by the $\Pi_i$ is of order $\hbar$. The effective dynamics 
produced by the full geometrical variational principle $\ref{gd}$
corresponds therefore to Hamiltonian dynamics.

 {\it Quantizing Free Dynamics on $T^\star\!{\cal M}$}:
The quantization of the extended dynamical system \ref{gd} proceeds
in a straightforward manner. It is infact equivalent to the quantization 
of a particle moving on a curved manifold in an external magnetic field.
Thought affected by ordering ambiguities arising from the non trivial
geometry the solution of this problem has been 
extensively discussed in the literature \cite{MONO}. 
 A little care has to be taken when the topology of the problem 
is non trivial. The Hilbert space of the system has to be constructed 
as that of square integrable sections of a line bundle $L$ over the 
`configuration space' ${\cal M}$ having the magnetic field $\omega_{ij}$ 
as curvature form.
 The same mathematical framework of pre-quantization 
is therefore recovered by means of the analogy of our theory with a 
magnetic system. As an example Kostant's quantization condition ensuring the 
existence of the line bundle $L$
$$ 
\int_\Sigma\omega=2\pi n,
$$ 
$\Sigma$ an arbitrary compact surface in ${\cal M}$ and $n$ an integer,
reappears as the Dirac's condition on monopole charge.
In any coordinate frame the quantum Hamiltonian describing the 
theory is given by
\begin{equation}
{\cal H}={1\over g^{1/2}}\Pi_ig^{ij}g^{1/2}\Pi_j +\hbar{\cal I}_1 
                                         +\hbar^2{\cal I}_2 +...
\label{hamq}
\end{equation}
where the kinematical momenta $\Pi_i=-i\hbar^{1/2}\partial_i-
\theta_i/\hbar^{1/2}$ have been introduced and ${\cal I}_1$,
${\cal I}_2$, ... are `optional' invariants reflecting the 
ordering ambiguities inherent the quantization procedure.
It is worthwhile to stress that ${\cal H}$ is a globally well
defined operator on the Hilbert space of the theory.

 Once quantization has been performed the same  mechanism
producing the reduction of the classical theory to Hamiltonian 
mechanics provides dynamically the reduction to the physical sector 
of the quantum theory. Depending only on the canonical formalism the 
argument goes exactly as the one in classical case and will not be repeated. 
By introducing  the guiding 
center operators $X^i=\xi^i+\hbar^{1/2}{\bar\omega}^{ij}\Pi_j$  
we obtain a set of operators fulfilling the canonical commutation relations
$[X^\mu,X^{\nu+n}]=i\hbar\delta^{\mu\nu}$ and $[\Pi_\mu,\Pi_{\nu+n}]=
-i\delta_{\mu\nu}$, $\mu,\nu=1,...,n$. Up to irrelevant terms the dynamics of
the physical variables $X^i$ separates from that of the $\Pi_i$
and Hamiltonian \ref{hamq} decomposes as in \ref{ham3}. 
 The energy necessary to induce a transition in the spectrum of 
the fast variables $\Pi$ being of order 1---to be compared with 
$\hbar$---the system behaves as frozen in one of the harmonic oscillator 
eigenstates of ${1\over2}\sum_i\Pi_i\Pi_i$ and dynamics is effectively 
reduced to the physical sector described by the $X$.

\end{document}